\documentclass[prc,aps,preprint,showpacs,nofootinbib]{revtex4}
\usepackage{graphicx}

\begin{document}
%\baselineskip 25pt minus .1pt
%\begin{center}
\title{A further look at prolate dominance in nuclear deformation} 
%[4ex]

\author{ Ikuko Hamamoto$^{1,2}$ and Ben R. Mottelson$^{2,3}$ }

\affiliation{
$^{1}$ {\it Division of Mathematical Physics, Lund Institute of Technology 
at the University of Lund, Lund, Sweden} \\ 
$^{2}$ {\it The Niels Bohr Institute, Blegdamsvej 17, 
Copenhagen \O,
DK-2100, Denmark} \\ 
$^{3}$ {\it NORDITA, Blegdamsvej 17, 
Copenhagen \O,
DK-2100, Denmark} \\ 
}

%\end{center}

%\vspace{2cm}

%\date{\today}

\begin{abstract}
The observed almost complete dominance of prolate over oblate deformations in
the ground states of deformed even-even nuclei is related to 
the splitting of high $\ell$ ''surface'' orbits in the Nilsson diagram: on the
oblate side the occurrence of numerous strongly avoided crossings which reduce
the fanning out of the low $\Lambda$ orbits, while on the
prolate side the same interactions increase the fanning out.
It is further demonstrated
that the prolate dominance is rather special for the restricted 
particle number of
available nuclei and is not generic for finite systems with mean-field
potentials resembling those in atomic nuclei.
\end{abstract}

%\vspace{1.5cm}

\pacs{21.60.Ev, 21.10.Pc, 21.90.+f}

\maketitle

%\newpage

\section{INTRODUCTION}
The ground states of some nuclei are spherical, while others are deformed
as a result of the latent anisotropy inherent in the ground state of 
a Fermi-gas.
In a very simplified view of the filling of particle orbits in the shell model
based on spherical symmetry, the ground state configurations at the beginning
and end of major shells are related by a particle-hole symmetry. 
Since the quadrupole moment of hole states has the opposite signs from that of
particle states one might have expected the number of prolate and oblate shapes
to be equal. 
In fact, almost all known deformed nuclei can be interpreted in terms of  
prolate axially-symmetric dominantly quadrupole deformed shape. 
The observed dominance of prolate over oblate shape is indeed overwhelming: 
Of the 98 known deformed even-even nuclei identified in Fig. 4.3 of 
Ref. \cite{BM75}, only one ($^{12}$C) is oblate.  Additional experimental work
since the review in Ref. \cite{BM75} has not provided evidence for more than a
few possibly oblate deformations.  
 
Calculating one-particle spectra of the quadrupole-deformed 
infinite-well potential 
(spheroidal cavity) which simulates the
potential for clusters of metallic atoms, we have obtained 
in Ref. \cite{HMXZ91}  
the result that the number of prolate systems is
considerably larger than that of oblate systems.
A number of publications using both quantum-mechanical and semi-classical 
treatments are available in which one tried to pin down the origin of 
the dominance of prolate nuclear shape over oblate shape.  For example, 
when Hartree-Fock (HF) 
calculations with appropriate effective interactions are performed 
in many nuclei, the
dominance of prolate shape except for very light nuclei is obtained in
agreement with the experimental observations.  
Nevertheless, in our opinion, the nature of the parameters responsible for the
prolate dominance has not yet been adequately understood.
In particular we are concerned with the generality of the
argument for prolate dominance: is it a universal property of saturating Fermi
systems bound by short-range interactions ?
It is the purpose of the present article to extend the calculations of Ref.
\cite{HMXZ91} with the aim of obtaining additional evidence concerning prolate
dominance.

In the present work 
only one kind of particles, neutrons or protons, are considered.
For simplicity, we neglect both spin-orbit potential and 
pairing correlation which are important in the understanding of nuclear ground
states, since an essential element in producing more prolate systems than 
oblate ones can be shown without them.  
In the spectra of both the harmonic oscillator potential and an isolated
single-j-shell (or single-$\ell$-shell) model 
there is a particle-hole symmetry, while in both of 
them the surface effect is absent.  The particle-hole
symmetry leads to the number of prolate systems equal to that of oblate ones. 
However, the presence of a more sharply defined surface leads to the presence of
''surface states'' that break the particle-hole symmetry. Even in the case of
the deformed harmonic oscillator, the deformations 
in mid-shell become so large that there occurs crossing between the one-particle
levels in
adjacent shells that violates particle-hole symmetry for the ground state
occupations.
We perform numerical calculations, choosing two models which can be 
easily solvable taking exactly 
into account the volume conservation when the system is quadrupole deformed; 
the pure harmonic oscillator potential and the infinite-well potential (cavity).
The numerical results of the two models are used to elucidate the important
role played by the surface of one-body potentials.

In Sec. II.A we study the quadrupole deformation preferred by the pure harmonic 
oscillator model, while in Sec. II.B the preference for 
prolate over oblate shape in
the spheroidal cavity is examined.
In Sec. III the origin of the dominance of prolate systems is explained using 
the numerical results presented in Sec. II.
In Sec. IV we present considerably extended numerical results of calculations of
ground states of independent particle motion in spheroidal cavities.
In Sec. V comments on some other approaches to the present
subject are briefly described, while conclusion and discussions are 
given in Sec. VI. 

\section{MODEL CALCULATION}

\subsection{Harmonic oscillator potential}
We parameterize the axially-symmetric quadrupole-deformed oscillator
potential by 
\begin{eqnarray}
\omega_{\perp} & = & \omega_{0} \, e^{\alpha} \\
\omega_{z} & = & \omega_{0} \, e^{-2\alpha}
\end{eqnarray}
so that the volume conservation
\begin{equation}
\omega_{\perp}^{2} \omega_{z} = \omega_{0}^{3}
\end{equation}
is exactly satisfied.
$\alpha > 0$ ($\alpha < 0$) corresponds to prolate (oblate) shape.  
For reference, in Table I a numerical comparison between $\alpha$, 
the ratio of radii $R_{z}/R_{\perp}$ (aspect ratio) 
and the commonly employed deformation parameter 
$\delta_{osc}$ \cite{BM75} is tabulated, where
\begin{equation}
\delta_{osc} \equiv 3 \, \frac{\omega_{\perp} - \omega_{z}}{2\omega_{\perp} +
\omega_{z}} = 3 \, \frac{R_{z} - R_{\perp}}{2R_{z} + R_{\perp}}
\end{equation}

The one-particle spectrum of the axially-symmetric quadrupole-deformed 
oscillator potential can
be found in many references.  See, for example, Fig. 6-48 of Ref. \cite{BM75} 
or Fig. 1 of Ref. \cite{SRN72}.  
Choosing the energy unit 
$\hbar \bar{\omega}$ where 
\begin{equation}
\bar{\omega} = \frac{1}{3} (2\omega_{\perp} + \omega_{z})
\end{equation}
one-particle energies can be drawn by straight lines as a function of
$\delta_{osc}$.
We note that a given level specified by quantum numbers ($n_{z}$, $n_{\perp}$) 
has a degeneracy 2($n_{\perp} + 1$) including the nucleon spin.
The magic numbers of the spherical harmonic oscillator potential are 
2, 8, 20, 40, 70, 112, 168, 240, ...
For a given nucleon (proton or neutron) number we numerically 
search for the total energy minimum on prolate and oblate sides, respectively. 
The total energy of the system in the present work is defined as the sum of 
the lowest-lying one-particle energies for a given deformation and 
a given particle number  
\begin{equation}
E(\alpha) = \sum_{i=1}^{N_{F}} \varepsilon_{\Lambda}^{i} (\alpha)
\label{eq:esum}
\end{equation}
where the conserved one-particle quantum-number $\Lambda$ is 
the projection of the particle orbital angular-momentum 
onto the symmetry axis. 

The one-particle spectrum originating from a given major $N$=$n_z + n_{\perp}$ 
shell 
has a symmetry between the prolate and oblate sides. 
Therefore, if for a given particle number we keep adiabatically 
the configuration consisting of 
the orbits occupied at a very small $| \delta_{osc} |$ value 
and look for the total energy minimum
as $|\delta_{osc}|$ further increases, the shape of the total 
energy minimum is prolate
for the Fermi level lying in the first half of the major shell 
while it is oblate for the second half of the major shell.  
Namely, the number of prolate systems is
equal to that of oblate systems.  This prolate-oblate symmetry may be broken, 
in the case that the energy minimum occurs at $|\alpha|$ values 
larger than
those at which one-particle levels coming from adjacent major shells cross
each other, namely in the case that the adiabatic
configuration defined 
above is no longer the configuration at the total energy 
minimum. 
In practice, this situation occurs only in a limited number of systems, 
in which the major shell is almost half filled.

In Fig. 1a the total energy of the prolate and oblate minima, respectively, 
relative to the energy of the 
spherical shape is plotted as a function of the particle number, of which 
the Fermi levels lie within the $N$=5 and 6 shells in the spherical limit. 
It is seen that in the
beginning (end) of the respective $N$ shells the prolate (oblate) minimum is 
deeper than the oblate (prolate) minimum, and as a total 
the number of prolate systems 
is approximately equal to that of oblate systems.  
In a few systems in which a given major shell is almost half filled, 
the lowest lying curves in Fig. 1a look somewhat irregular, however, 
the curves become much 
smoother when a deviation from axial symmetry is taken into account.  
Similarly, in the region of the 
particle number for which the curve in Fig. 1a locally 
resembles a straight line, taking into account a deviation from axial symmetry 
makes the curve more smoothly varying.  
In all these cases, the deviation from respective
axially-symmetric deformations is relatively small. 
 
In Fig. 1b absolute values of $\alpha$ at the total energy minimum for a given
particle number are plotted, except 
around the middle of respective major 
shells in which 
the $| \alpha |$ values of both prolate and oblate minima are shown.
In the latter cases, 
precisely speaking for the particle number 84-90 and 134-136, 
the energy minimum is obtained at $|\alpha|$ which is larger than
that of the crossing of one-particle levels coming from adjacent major shells. 
The presence of those few
systems provides a small deviation from the exact equality of the number 
of prolate and oblate shapes 
in the harmonic oscillator potential.

\subsection{infinite-well potential (cavity)}
The eigenvalue of spherical cavity $\varepsilon_{n \ell}$ is obtained from the 
$n$th zero of the spherical Bessel function of order $\ell$.
In Fig. 2a eigenvalues of relatively small spherical cavities 
together with the total particle number $A$ for
several Fermi levels are plotted as a function of orbital angular
momentum $\ell$, while in Fig. 2b eigenvalues of 
a much broader region are shown.

From Fig. 2b two kinds of families of one-particle levels ($n \ell$) are
easily identified; (a) a family
with $\Delta \ell$=2 
originating from a given harmonic-oscillator major shell,  
of which the one-particle energies decrease as $\ell$ increases. 
This decrease of one-particle energies in realistic potentials is
approximated by introducing 
the $\vec{\ell}^2$ term in the modified oscillator potential
\cite{SGN55}; 
(b) families defined by the number of radial nodes $n$=0, 1, 2, ... and 
each comprising all possible $\ell$-values $\ell$=0, 1, 2, ...  
These families are referred to as $n$=0=yrast, $n$=1=yrare, ...

It is seen that in the family (a) the $\Delta \ell$=2 approximate degeneracy
around $\ell$=0 remains all the way to very large systems, 
corresponding
to elliptical orbits in terms of closed classical orbits \cite{BM75}.
Since the total degeneracy of those $\Delta \ell$=2 close-lying levels  
in the family is small compared with the degeneracy 
of all possible open shells around the Fermi levels,
those levels do not play an important role in the present discussion of the
prolate-oblate competition.  In contrast, orbits close to the yrast line have
the largest $\ell$ values and thus the largest degeneracies among orbits around a
given Fermi level.  Thus, the splitting or the shell structure of those large
$\ell$ shells for quadrupole deformation may govern the preferred prolate or
oblate shape.

As seen in Fig. 2a, for 
138$<A<$186 almost degenerate 1j and 2g shells are considerably separated
from other $n \ell$ shells.  The near degeneracy of these two
shells with $\Delta \ell$=3 may lead to octupole deformations.  
In Ref.
\cite{HMXZ91} this possibility is examined under the assumption of deformations
involving only a single spherical harmonic mode.  It is found that quadrupole
mode dominates except for the particle number
$A$=152-156 where Y$_{32}$
deformation provides the lowest total energy.

The radii of the spheroidal cavity are parameterized as 
\begin{eqnarray}
R_{\perp} & = & R_{0} \, e^{- \alpha} \\
R_{z} & = & R_{0} \, e^{2 \alpha}
\end{eqnarray}
so that the volume conservation is exactly satisfied under deformation.
The numerical 
relation between $\alpha$ and $R_{z}/R_{\perp}$ is given in Table I.
Eigenvalues of spheroidal cavity are calculated using the method described by
S. A. Moszkowski \cite{SM55}.  
The Nilsson diagram for the spheroidal cavity is shown in Figs. 3a and 3b
, in which one-particle levels with $\Lambda$=0 are doubly degenerate while
those with $\Lambda \neq 0$ are four-fold degenerate.
It is noted that Figs. 3a and 3b cover a region of the deformation which is as 
much as a factor of two 
larger than that of possible ground states.  See Fig. 4b.  

In Fig. 4a the total energies at the minima of the prolate and oblate shapes,
respectively, relative to the
energy of the spherical shape are plotted for systems with 
even particle-number 92-186. 
What is shown in Fig. 4a is indeed identical to a part of Fig. 23 of Ref. 
\cite{HMXZ91}.  
As already stated in Ref. \cite{HMXZ91}, we have found : (a) at the beginning of
the two major shells (the particle number 92-138 and 
138-186) the optimum 
oblate shape is energetically more favorable than the optimum prolate shape 
while at the end of the
shells the optimum prolate shape is more favorable; 
(b) the number of systems in which a 
prolate shape is favorable is much larger than the
number for which an oblate shape has a lower energy.  For example,
for the major shell with the particle number 138-186 
only six systems prefer oblate while the prolate minimum is lower 
than the oblate minimum in seventeen systems. 

Absolute values of $\alpha$ at the energy minima of prolate and
oblate shapes, respectively, 
are plotted in Fig.4b.
It is seen that in the mid-shell region the absolute magnitudes of the
deformations are appreciably larger for the prolate than for the oblate
deformations.

\section{Origin of the dominance of prolate systems}
The origin of the dominance of prolate systems obtained in the spheroidal 
cavity comes from an asymmetry in the splitting of one-particle 
levels on the 
prolate and oblate sides, which is absent in the axially-symmetric 
quadrupole-deformed harmonic
oscillator model. The asymmetry originates from the fact that already in the 
spherical shape the presence of the surface in the potential 
implies that the higher $\ell$ sub-shells have lower energies than the lower
$\ell$ orbits of the same oscillator shell.
When the potential is moderately deformed, around the Fermi level of the
system where high $\ell$ shells are partially filled 
the local one-particle level density 
is considerably higher on the oblate side than on the prolate side.  
The origin of the different splittings of one-particle levels 
coming from high $\ell$ shells on prolate and oblate sides 
can be understood in terms of  
the asymptotic quantum numbers which have been found useful in the deformed 
oscillator model.

As an example of the splitting of the yrast family orbits, 
in Fig. 5 we reproduce 
the one-particle levels originating from the 1$h$ shell which are taken 
from Fig. 3b. 
It is noted that the splitting of 
$\ell$ orbits belonging to the yrast family in Fig. 2b  
is all very similar.    
Two characteristic features in the level splitting are noted;
(a) on the oblate side strongly avoided crossings among the low $\Lambda$ 
orbits with resulting reduction in fanning out; (b) on the prolate side 
increased fanning out  
of the low $\Lambda$ levels due to the same inter-shell interactions.

In Fig. 5 the quantum numbers [N n$_z$ $\Lambda$],
which are called the asymptotic quantum numbers 
in the deformed oscillator model \cite{BM75}, are assigned to respective 
one-particle 
levels, where $\Lambda$ is a good quantum number also in the spheroidal cavity.
The quantum number n$_z$ in the deformed oscillator potential 
is usually known 
as the number of oscillator quanta in the direction of
the symmetry axis (z-axis).  Generally speaking, for a prolate (oblate) shape 
the kinetic energy is energetically 
cheaper (more expensive) in the direction of 
the symmetry axis.
In the spheroidal infinite-well potential $n_z$ can be interpreted as 
the number of the node of the wave function in
the direction of the symmetry axis, while $N$ represents
the sum of the nodes of the wave functions in the directions of the symmetry
axis and the x and y axes.  
Defining the meaning of the quantum numbers, $N$ and n$_z$, in this way, 
one obtains the asymptotic behavior of one-particle levels in 
the spheroidal cavity as follows; 
(i) For a large deformation the quantum numbers [N n$_z$ $\Lambda$] 
become good quantum 
numbers; (ii) For a large deformation the slope of the one-particle energies  
in Figs. 3a, 3b and 5 is determined by the quantum numbers N and n$_z$; 
(iii) Levels with
larger n$_z$ for a given N lie energetically 
lower (higher) in prolate (oblate) shape; (iv) The presence
of the surface in the potential makes 
the levels with larger $\Lambda$ values for given N and n$_z$ lower, 
since already in the
spherical shape higher $\ell$ shells are energetically pushed
down compared with lower $\ell$ shells.  

On the prolate side (i.e. $\alpha > 0$) 
of Fig. 5 the levels split for a small $\alpha$ value
have already the internal structure which smoothly changes to respective  
asymptotic quantum numbers 
and, thus, the splitting grows
smoothly and monotonically as $\alpha$ further increases.  In contrast, 
on the oblate side (i.e. $\alpha < 0$) the levels except two levels with [505] 
and [514] have to change drastically the internal 
structure to approach their asymptotic quantum numbers as
$|\alpha|$ increases.
The drastic change of the internal structure 
comes from the interaction (or the avoided crossing) 
with the one-particle levels
originating from the 2f and 3p shells.  Consequently, it produces 
the strongly non-linear behavior for
those one-particle levels on the oblate side, in striking contrast to the
prolate side, in the deformation region relevant to that of the ground states.
More generally speaking, for a given $N \gg 1$ the splitting of 
one-particle levels coming from the highest $\ell$ (= $N$) sub-shell 
grows monotonically on the prolate side, while on the oblate side all levels
except the two levels with [N 0 N] and [N 1 N-1] have to change drastically 
the internal
structure very soon after $| \alpha |$ increases from zero. 
From Fig. 3b one can indeed see 
that the splitting structure of one-particle levels originating from the 1i and
1j shells is very similar to Fig. 5 on both prolate and oblate sides.

The similar asymmetry of the splitting of one-particle levels on  
the prolate and oblate
sides can be identified also for
one-particle levels coming from the $\ell$ = N$-$2 shell (the yrare family in
Fig. 2b) if $N$ is 
sufficiently large, though the asymmetry is less striking.
If we take an example of the one-particle levels coming from the $N$=5, 
2f shell 
in Fig. 3b, the one-particle levels on the prolate side have the asymptotic
quantum numbers, [530], [521], [512] and [503], from the bottom to the top,
and the level splitting grows monotonically as $\alpha$ increases from zero.  
In contrast, on the oblate side 
the asymptotic quantum numbers are [523], [521], [532] and [530], from the
bottom to the top.  That means, two levels starting with [512] and [503] at 
very small values of $|\alpha|$ on the oblate
side have to change
the internal structure soon after $| \alpha |$ increases from zero.  
Consequently, 
in the deformation region relevant to the ground states 
the local one-particle level density 
on the oblate side is higher than that on the prolate side.
This helps further the prolate dominance produced by the level splitting of 
the 1i ($\ell$=N=6) shell in the region of 
particle number 94-130.

The level splitting of the $\ell \ll N$ orbits does not play an important role 
in the present discussion, since the
degeneracy of these sub-shells is small. 

For reference, in Fig. 6 the level splitting of an isolated single $\ell$=5 
shell in cavity is shown, which is obtained by switching off the coupling to
other shells.  The volume conservation taken into account is the same as the one
in Fig. 5 and leads to the curves in Fig. 6 instead of straight lines
which may often be seen in the literatures as the splitting of 
an isolated single $\ell$ shell.  
In the case of this isolated single $\ell$ shell 
the wave functions of all one-particle levels are independent of deformation, 
and there is a particle-hole symmetry or a symmetry
between the prolate and oblate sides. The shape of the total 
energy minimum is oblate
for the Fermi level lying in the first half of the shell while it is prolate in 
the second half of the shell.  The preference for prolate or oblate shape in
the beginning and the end of the shell, respectively, is 
opposite to that in a major shell of the pure harmonic oscillator potential.

From the comparison of the splitting on the prolate side 
of Fig. 6 with that of Fig. 5,
it is noted that the interaction with the normal-parity states in the next shell
above implies non-linear convexity in the lower $\Lambda$ orbits in Fig. 5. 
This increase of fanning out is another
important factor favoring prolate dominance.

So far, using the cavity potential 
we have explained the dominance of prolate shape  
in terms of the prolate-oblate 
difference of the bunching of high-$\ell$ ''surface'' states, which are
recognized as a surface mode bound to surface by the large 
centrifugal potential.  
The characteristic feature of the bunching of those high $\ell$ 
one-particle levels, 
can be found in all realistic nuclear potentials.  It comes from 
the presence of the surface in the Woods-Saxon potential and Hartree-Fock
potentials and is parameterized by the $\ell ^2$ term in the modified
oscillator potential.
The bunching unique to the prolate and oblate shapes, respectively, 
described above can be found already 
in the Nilsson diagram of a small system such as 
the sd-shell in realistic nuclear potentials. 
For example, see the splitting of the levels originating from 
the 1d$_{5/2}$ shell 
in Fig. 5-1 of Ref. \cite{BM75}, where the additional effect of the
spin-orbit potential included should be also noticed.

\section{Prolate-oblate competition in spheroidal cavity for larger systems}
In previous sections 
we have shown that the different bunching of Nilsson levels for cavity on
the prolate and oblate sides, which come from high $\ell$ shells,
leads to the prolate dominance in deformed nuclei. 
While the number of particles which can be accommodated 
in the highest $\ell$ shell (the yrast family in Fig. 2b) 
is a large portion of the particle number
accommodated in one 
major shell of smaller systems, 
the portion becomes smaller in a larger system.
This is because the number of particles accommodated in the highest $\ell$ shell
in a spherical potential is the order of A$^{1/3}$ where the total number of
particles is expressed by A, while the degeneracy of one major shell is the
order of A$^{2/3}$ (A$^{1/2}$) in the harmonic oscillator potential (in 
potentials such as an infinite-well potential). 

The calculation of Ref. \cite{HMXZ91} of equilibrium shapes of independent
particles in a spheroidal cavity have been extended to particle number 
up to 850 with a view to examine the prolate-oblate competition 
in a more general context\footnote{Motivated by the issue of deformations 
in the sodium clusters,
similar calculations based on the shell correction method without a detailed
discussion of 
prolate-oblate competition have been carried out by Reimann and Brack in 
Ref. \cite{RB94}.}.

In Fig. 7 we show 
eigenvalues of spherical cavity for systems larger than those in
Fig. 2a and up till those investigated in Ref. \cite{RB94}. 
As some numerical examples, 
in Figs. 8a and 9a the total energies at the energy minima relative to the
energies of the spherical shape as a function of particle number are plotted for
systems with the particle number 338-440 and 676-832, respectively. 
In Figs. 8b and 9b the absolute values of $\alpha$ at the total energy minima 
are shown.  
The close relation between the prolate or oblate minima and the location of 
high-$\ell$ orbits can be easily seen in the same way as that in the
smaller systems described previously.
In Table II we tabulate the calculated number of prolate and oblate systems, 
and the 
ratio of the number of oblate system to that of
all deformed (namely oblate plus prolate) systems.
From Table II it is seen that the dominance of prolate shape over oblate shape
is gradually reduced as the particle number increases. 
The decrease of deformation with particle number shown in Table II follows
rather well the expected power law $A^{-1/2}$.

\section{Comments on some other approaches}
Exploiting the theory of periodic orbits, 
H. Frisk \cite{HF90} has suggested that
the dominance of prolate systems over oblate systems originates in the landscape
of the locally averaged one-particle level density considered 
as a 
function of particle number and spheroidal deformation parameters 
in the potential.
It is however 
difficult for us to assess the scope of the approach in Ref. \cite{HF90} 
because it fails to provide quantitative measures for the relative number of
prolate and oblate deformations.

A variety of theoretical models, some of which 
are fully microscopic while others are 
some combinations of macroscopic (or phenomenological) and microscopic 
approaches, have been used in the study of the prolate-oblate competition 
of the ground states of deformed nuclei.
Here, as an example, we take the work by N.Tajima et al. in Ref. \cite{NT96}, 
which is an HF plus BCS calculation with the SIII interaction 
as the HF effective interaction.
The ground states of even-even nuclei with the proton number 2$\leq$Z$\leq$114 
and the neutron number N ranging from outside the proton drip line to beyond 
the experimental frontier on the neutron-rich side are considered.
In Ref. \cite{NT96} it is found that in heavier nuclei 
the oblate ground states are very rare.  They state that the dominance of
prolate deformation for N$>$50 may be attributed to the change of the nature of
the major shells from the harmonic-oscillator shell to the Mayer-Jensen shell.
This statement can be interpreted as : 
the spin-orbit splitting is an essential
element in the dominance of prolate systems.  
In order to clarify the role by the spin-orbit potential in the 
prolate-oblate competition, let us briefly consider a model consisting of 
the harmonic oscillator plus spin-orbit potentials.  In the spherical
limit of this simple model 
the splitting of one-particle energies in a given major $N$ shell comes 
only from the spin-orbit potential.  
As an example, we take the $N$=5 major shell.  Then, taking the sign of the
spin-orbit potential in nuclear physics, in the spherical limit  
the lowest-lying shell among j shells 
belonging to the major shell is 1h$_{11/2}$, 
while the highest is the 1h$_{9/2}$ shell. 
It is easy to find that in both prolate and oblate sides 
the level splitting of the 1h$_{11/2}$ shell is 
similar to that shown in Fig.5, when $\Lambda$ is replaced by $\Omega =
\Lambda + s_{z}$ where $\Omega$ exåresses the nucleon angular-momentum
component along the symmetry axis. 
In contrast, the splitting of the levels of the highest-lying 
1h$_{9/2}$ shell is
quite different; On the oblate side the splitting for a small $|\alpha|$ grows
smoothly and monotonically as $|\alpha|$ further increases (just like the level
splitting on the prolate side of Fig. 5), while on the prolate side
all one-particle levels must have asymptotically $N$=5 and $n_{z}$= 0 or 1.  
Namely, all levels originating from the 1h$_{9/2}$ shell 
on the prolate side must 
be asymptotically up-sloping as $\alpha$ increases.  
That means, on the prolate side 
three levels other than the two levels, [505 9/2] and [514 7/2], 
have to change the internal structure from that 
for very small $\alpha$ values as $\alpha$ further increases, 
in order to approach the asymptotic behavior. 
Thus,  
we expect more prolate systems for the Fermi level lying at the beginning of the
N=5 major shell (due to the levels coming from the 1h$_{11/2}$ shell), 
while more oblate systems may be obtained 
for the Fermi
level lying at the end of the major shell (due to the levels coming from the
1h$_{9/2}$ shell).  
As a total, we conclude that 
the spin-orbit potential alone may not have any strong preference
for prolate shape over oblate shape.

\section{conclusion and discussions}
This article began with the question of why almost all deformed nuclei are
prolate in their ground state rather than an equal division between prolate and
oblate as suggested by the particle-hole symmetry that follows if we ignore
sub-shell structure of spherical major shells.  We have only been able to
identify a mechanism that leads to a significant dominance of prolate over
oblate shapes in the range of particle number up to those 
experimentally examined.
This mechanism involves the transition from harmonic oscillator mean field to
systems with a much more sharply defined surface, modeled by a deformed cavity.
In the latter region there occur surface states with orbital angular momentum
and degeneracy appreciably greater than any other in the shell.  This pattern
leads to the occurrence of avoided crossings on the oblate side of the Nilsson
diagram.  This leads to shell filling in which the deformation is oblate in the
beginning of the shell, but changes to prolate well before mid-shell because 
of the energy gain implied by the fanning out of the prolate orbits 
that is increased by the interaction with the normal-parity sub-shells in the
next higher major shell.
The role played by the surface becomes less important for much larger systems.  
Thus, the observed overwhelming dominance of prolate shape in deformed nuclei
may be identified as the feature of the system 
with a relatively small number of particles.

In the present work we focus our attention on the prolate shape dominance 
in the ground states of stable or well-bound nuclei.  
In drip line nuclei with weakly bound 
nucleons, especially neutrons, the shell structure around the Fermi level 
as well as the
role played by the nuclear surface can be different \cite{HM03}.   
Therefore, the
prolate-oblate competition in drip line nuclei has to be separately and 
more carefully examined.

We recognize that the origin of the prolate dominance presented in this work 
is the essential element but it 
does not immediately lead to the observed overwhelming prolate 
dominance in the ground
states of deformed nuclei.  
Thus, in the following we make comments on some elements which have not been 
taken into account in our present work. 
First, we have not included the spin-orbit potential, 
which is important in nuclear spectroscopy.
We conclude that already in the absence of the
spin-orbit potential 
the dominance of prolate systems is obtained and the spin-orbit potential alone 
has no strong
preference for prolate or oblate systems.
However, it is possible that a particular 
combination of the surface effect of the nuclear
potential with the spin-orbit splitting may make a further contribution 
to the prolate 
dominance.  In particular, N. Tajima and N. Suzuki in Ref. \cite{TS01} have
identified an interesting coherence in the contributions of the $\ell^{2}$ and 
the spin-orbit terms to the prolate/oblate competition.  It is noted that
the spin-orbit potential is closely related to the surface property of
systems.

Second, we have not taken into account the pairing correlation. 
It is reported in Ref. \cite{TSS02} that 
the inclusion of the pairing correlation 
may enhance the prolate dominance, depending on the pairing strength. 
As seen in Figs. 4a and 4b, the systems with oblate minima occur in the
neighborhood of closed shells and the deformation is relatively small. 
Therefore, some of those oblate systems may easily become spherical when 
pair correlation is included.  Then, the ratio of prolate systems to oblate
ones may become larger, in agreement with the numerical results of 
Ref. \cite{TSS02}.

Third, 
the Coulomb interaction between protons clearly prefers prolate shape 
to oblate shape as exhibited by the cubic term, $\alpha^3$, in the Coulomb
energy of a deformed uniformly-charged ellipsoid.  
However, for the moderate deformation such as that of nuclear ground states the
preference is expected to play a minor role.

Fourthly, only one kind of nucleons (protons or neutrons) are considered in our
present work. 
Different shapes may be preferred by protons and neutrons in some nuclei 
with N$\neq$Z, and the possible difference may make a minor modification of 
the degree of the dominance of prolate systems.

%\end{references}

\newpage

%%%%%%% TABLE 1 %%%%%%%%%%%%%%%%%%%%%%%%%%%%%%%%%%%%%%%%%%%%
\begin{table}[t]
\caption{\label{tab:table1} Relation between parameters}
\vspace{2pt}

\begin{tabular}{|c|c|c|c|c|c|c|c|} \hline

$\alpha$ & -0.15 & -0.10 & -0.05 & 0 & 0.05 & 0.10 & 0.15 \\ \hline
$R_{z}/R_{\perp}$ & 0.638 & 0.741 & 0.861 & 1.000 & 1.162 & 1.349 & 1.568 \\ 
\hline
$\delta_{osc}$ & -0.477 & -0.313 & -0.153 & 0 & 0.146 & 0.283 & 0.412 \\

\hline

\end{tabular}

\end{table}

%\mbox{}

\vspace{3cm}

%%%%%%% TABLE 2 %%%%%%%%%%%%%%%%%%%%%%%%%%%%%%%%%%%%%%%%%%%%
\begin{table}
\caption{\label{tab:table2} 
The number of the prolate- and oblate-deformed systems 
and the maximum deformation 
for an infinite-well potential (spheroidal
cavity). 
The first column shows the region of the particle number, of which only the
system with even particle-number is examined.  The second column denotes the
number of prolate-deformed system, while the number of oblate-deformed system is
given in the third column. The ratio of the number of oblate-deformed system to
the sum of the prolate- and oblate-deformed systems is shown in the fourth
column, while the $\alpha$ value of the maximum $|\alpha|$ in respective
regions of the particle number is given in the fifth column.}
\vspace{2pt}
%\begin{center}
\begin{tabular}{c|c|c|c|c} \hline
  particle number & prolate & oblate &  
  ratio of oblate to total  & $\alpha$ of $\mid \alpha_{max} \mid$ \\ \hline
 58-92  & 11 & 4 & 0.27 & 0.081 \\ 
 
 92-138 & 15 & 6 & 0.29 & 0.072 \\
 
 138-186 & 17 & 6 & 0.26 & 0.059 \\
 
 186-254 & 24 & 7 & 0.23 & 0.064 \\
 
 254-338 & 28 & 13 & 0.32 & 0.060 \\ 
 
 338-440 & 38 & 11 & 0.22 & 0.050 \\
 
 440-556 & 35 & 20 & 0.36 & 0.043 \\
 
 556-676 & 38 & 19 & 0.33 & $-$0.029 \\
 
 676-832 & 45 & 29 & 0.39 & $-$0.027 \\ \hline
\end{tabular}
%\end{center}
\end{table}

\mbox{}

\newpage

{\bf\large Figure captions}\\
%%%%%%% FIG 1a %%%%%%%%%%%%%%%%%%%%%%%%%%%%%%%%%%%%%%%
%\begin{figure}
\begin{description}
%\includegraphics{fig1}
%\caption{\label{fig:fig1} 
\item[{\rm Figure 1a :}]
Total energies at the prolate and oblate minima, respectively,  
relative to the energy of the spherical shape as a function of  
particle number.  The harmonic oscillator potential is used and 
the energy unit is $\hbar \omega_{0}$.
\end{description}
%\end{figure}

%%%%%%% FIG 1b %%%%%%%%%%%%%%%%%%%%%%%%%%%%%%%%%%%%%%%
\begin{description}
\item[{\rm Figure 1b :}]
Absolute values of the deformation parameter $\alpha$ at the total 
energy minimum as a function of particle number.  
Around the middle of respective 
major shells the $| \alpha |$ values of both prolate
and oblate minima are shown.
\end{description}

%%%%%%% FIG 2a %%%%%%%%%%%%%%%%%%%%%%%%%%%%%%%%%%%%%%%
\begin{description}
\item[{\rm Figure 2a :}] 
Eigenvalues of smaller spherical cavity in units of $\hbar^2 / 2mR_0^2$ 
as a function of one-particle orbital angular
momentum. The total 
particle number including the
factor 2 due to the nucleon spin is shown for several Fermi levels.
\end{description}

%%%%%%% FIG 2b %%%%%%%%%%%%%%%%%%%%%%%%%%%%%%%%%%%%%%%
\begin{description}
\item[{\rm Figure 2b :}]
Eigenvalues of spherical cavity of a region much broader 
than that plotted in Fig. 2a.  An example of the family (a) coming from a given
harmonic-oscillator major shell is denoted by a dotted curve, while the yrast
$\Delta \ell$=1 family (b) with no radial node is connected by a dashed curve. 
\end{description}

%%%%%%% FIG 3a %%%%%%%%%%%%%%%%%%%%%%%%%%%%%%%%%%%%%%%
\begin{description}
\item[{\rm Figure 3a :}]
One-particle energies of spheroidal cavity 
as a function of deformation parameter.
At spherical point $\alpha$=0 the quantum numbers, $n \ell$, are written.
The particle number of the system obtained by filling all lower-lying 
levels is written with a circle in several places. 
Positive-parity levels are plotted by solid curves, while negative-parity 
levels by dotted curves.
\end{description}

%%%%%%% FIG 3b %%%%%%%%%%%%%%%%%%%%%%%%%%%%%%%%%%%%%%%
\begin{description}
\item[{\rm Figure 3b :}]
One-particle energies of spheroidal cavity as a function of deformation
parameter, for the system larger than that plotted in Fig. 3a.
See the caption to Fig. 3a.
\end{description}

%%%%%%% FIG 4a %%%%%%%%%%%%%%%%%%%%%%%%%%%%%%%%%%%%%%%
\begin{description}
\item[{\rm Figure 4a :}]
Total energies at the prolate and oblate minima, respectively,  
relative to the energy of the spherical shape as a function of  
particle number of the system.  The infinite-well potential is used and the
energy unit is $\hbar^2 / 2mR_0^2$. 
\end{description}

%%%%%%% FIG 4b %%%%%%%%%%%%%%%%%%%%%%%%%%%%%%%%%%%%%%%
\begin{description}
\item[{\rm Figure 4b :}]
Absolute values of the deformation parameter $\alpha$ at the  
energy minima of prolate and oblate shapes, respectively.  
\end{description}

%%%%%%% FIG 5 %%%%%%%%%%%%%%%%%%%%%%%%%%%%%%%%%%%%%%%
\begin{description}
\item[{\rm Figure 5 :}]
Splitting of levels originating from the 1h shell in spheroidal cavity.
The asymptotic quantum numbers [N n$_z$ $\Lambda$] 
are assigned to the levels on both prolate and
oblate sides.  See the text for details.
\end{description}

%%%%%%% FIG 6 %%%%%%%%%%%%%%%%%%%%%%%%%%%%%%%%%%%%%%%
\begin{description}
\item[{\rm Figure 6 :}]
Splitting of levels coming from an isolated 1h shell, which is obtained by 
switching off the coupling to other shells in spheroidal cavity. 
\end{description}

%%%%%%% FIG 7 %%%%%%%%%%%%%%%%%%%%%%%%%%%%%%%%%%%%%%%
\begin{description}
\item[{\rm Figure 7 :}] 
Eigenvalues of spherical cavity, which are larger than those 
in Fig. 2a, in units of $\hbar^2 / 2mR_0^2$ 
as a function of one-particle orbital angular
momentum. The total 
particle number including the
factor 2 due to the spin 1/2 is shown for several Fermi levels.
\end{description}

%%%%%%% FIG 8a %%%%%%%%%%%%%%%%%%%%%%%%%%%%%%%%%%%%%%%
\begin{description}
\item[{\rm Figure 8a :}]
Total energies at the energy minima  
relative to the energy of the spherical shape as a function of 
particle number 338-440 of the system.  
The infinite-well potential is used and the
energy unit is $\hbar^2 / 2mR_0^2$. 
\end{description}

%%%%%%% FIG 8b %%%%%%%%%%%%%%%%%%%%%%%%%%%%%%%%%%%%%%%
\begin{description}
\item[{\rm Figure 8b :}]
Absolute values of the deformation parameter $\alpha$ at the  
energy minima of Fig. 8a.  
\end{description}

%%%%%%% FIG 9a %%%%%%%%%%%%%%%%%%%%%%%%%%%%%%%%%%%%%%%
\begin{description}
\item[{\rm Figure 9a :}]
Total energies at the energy minima  
relative to the energy of the spherical shape as a function of  
particle number 676-832 of the system.  
The infinite-well potential is used and the
energy unit is $\hbar^2 / 2mR_0^2$. 
\end{description}

%%%%%%% FIG 9b %%%%%%%%%%%%%%%%%%%%%%%%%%%%%%%%%%%%%%%
\begin{description}
\item[{\rm Figure 9b :}]
Absolute values of the deformation parameter $\alpha$ at the  
energy minima of Fig. 9a.  
\end{description}

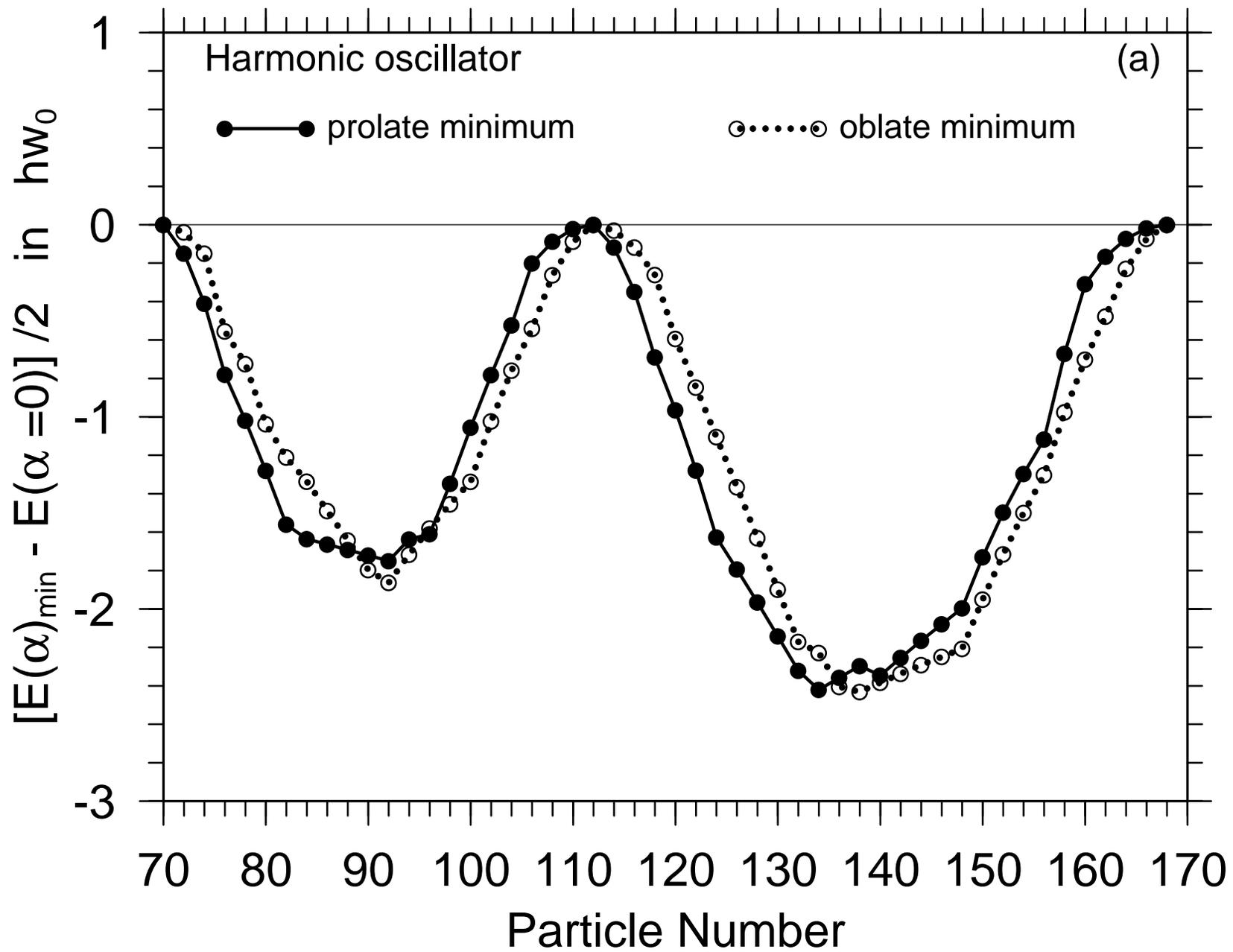
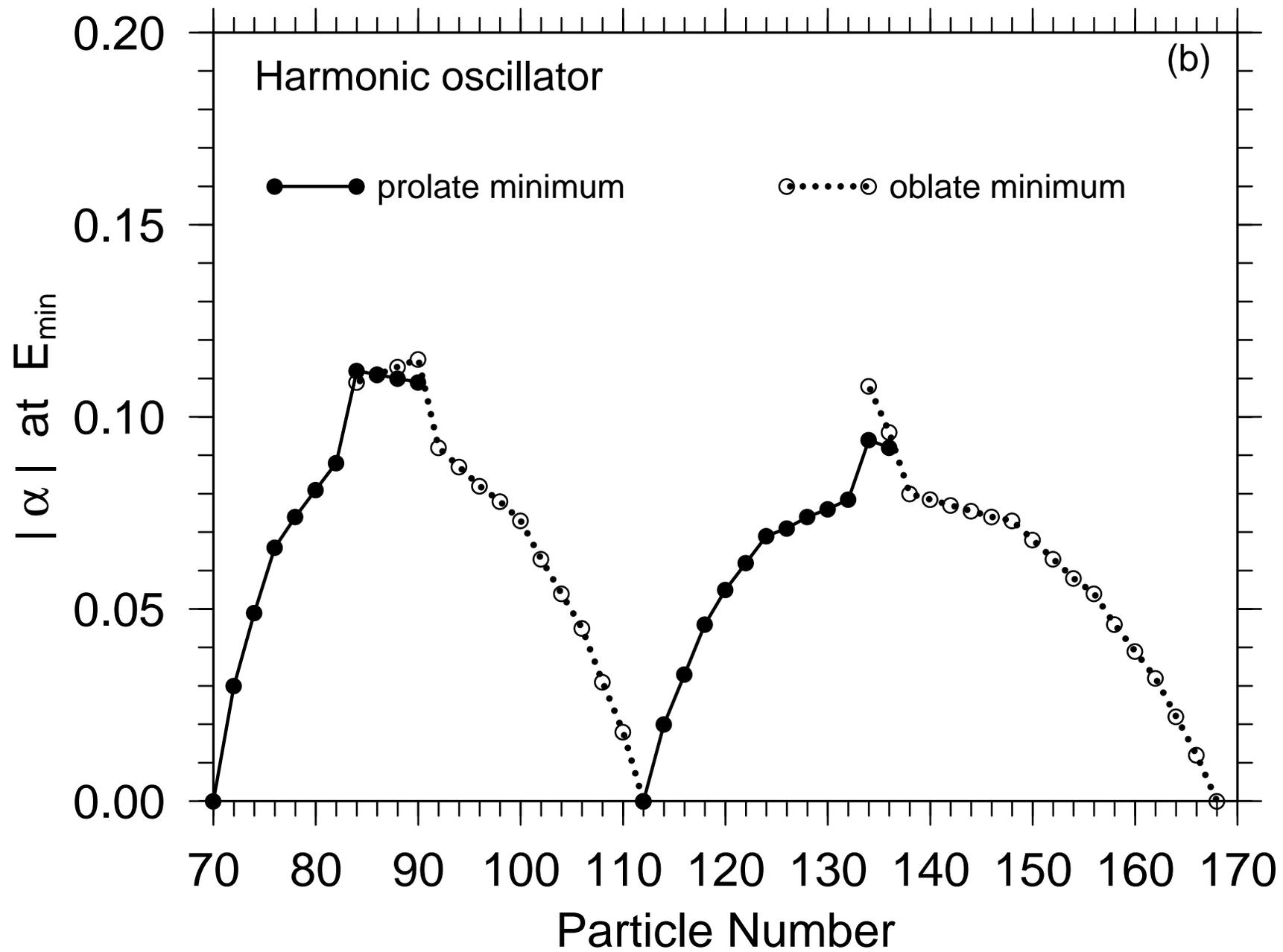
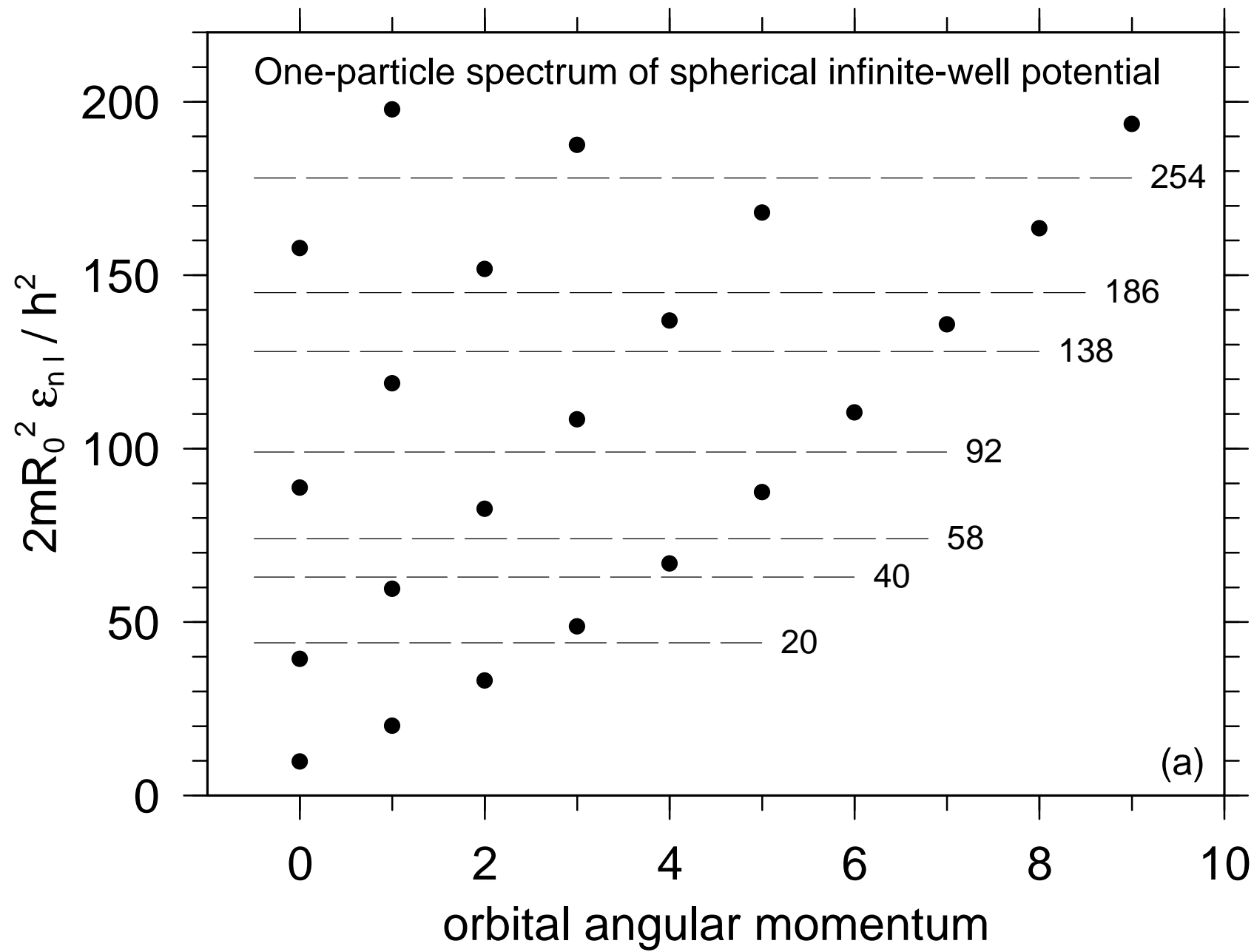
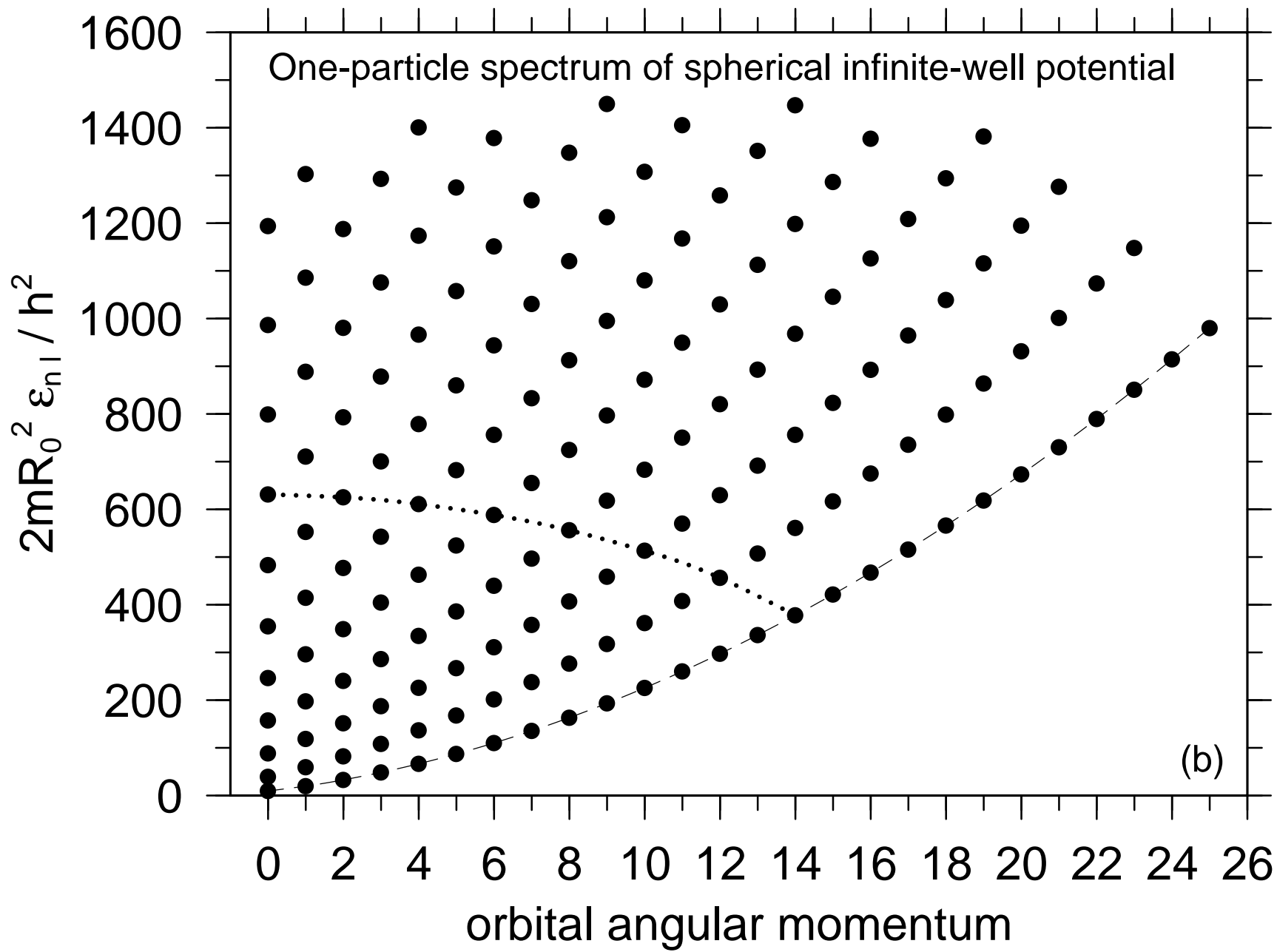
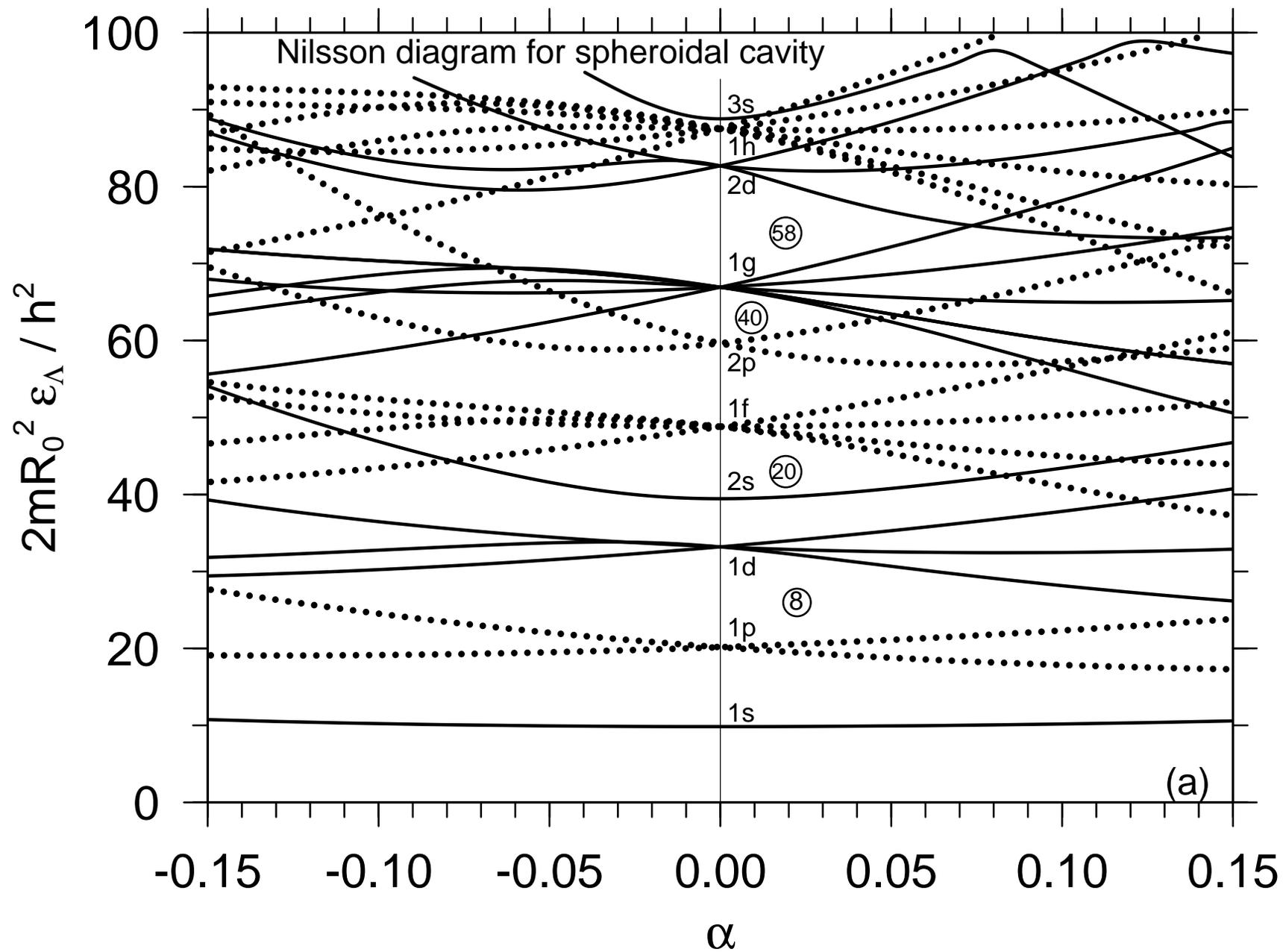
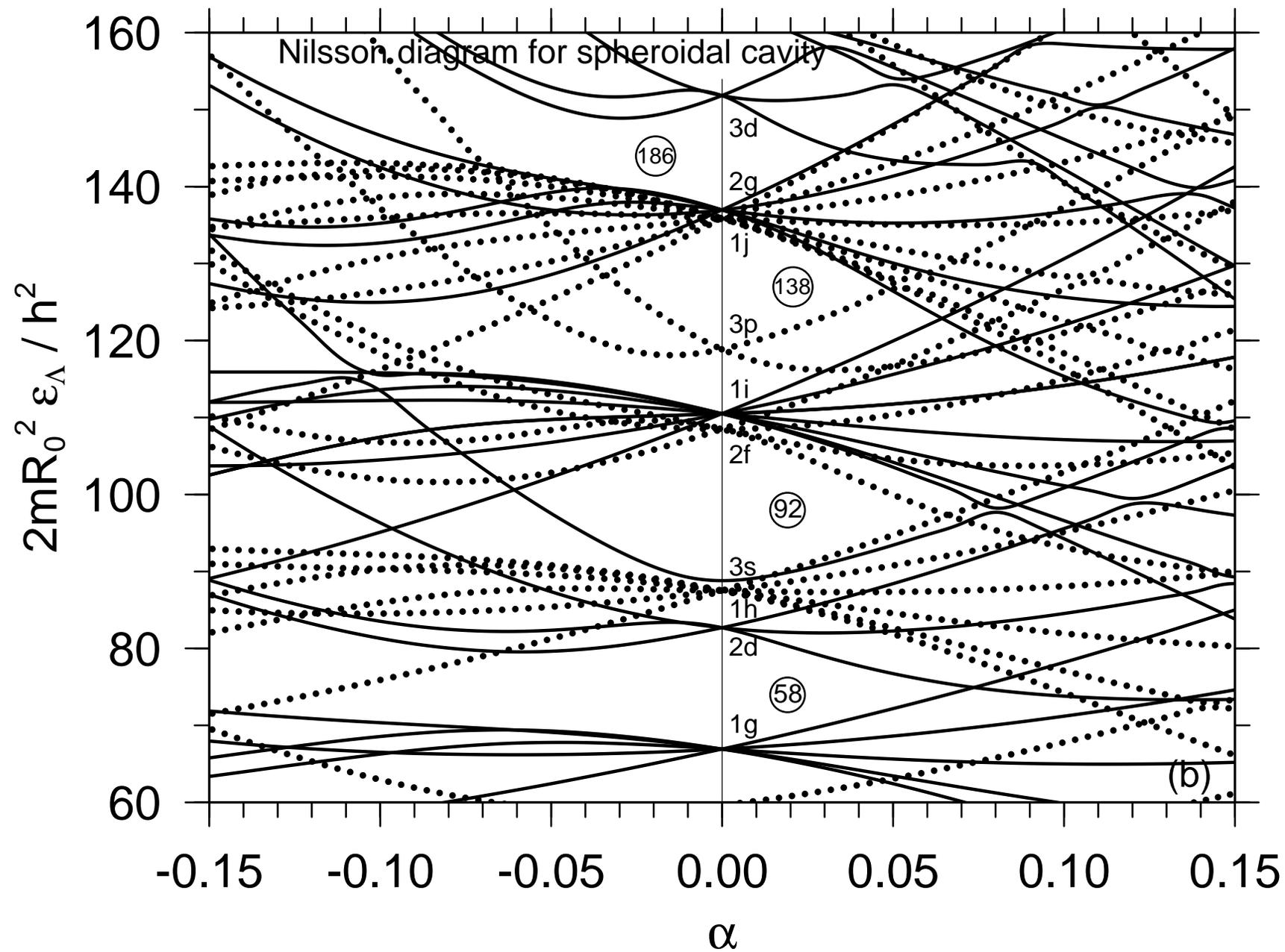
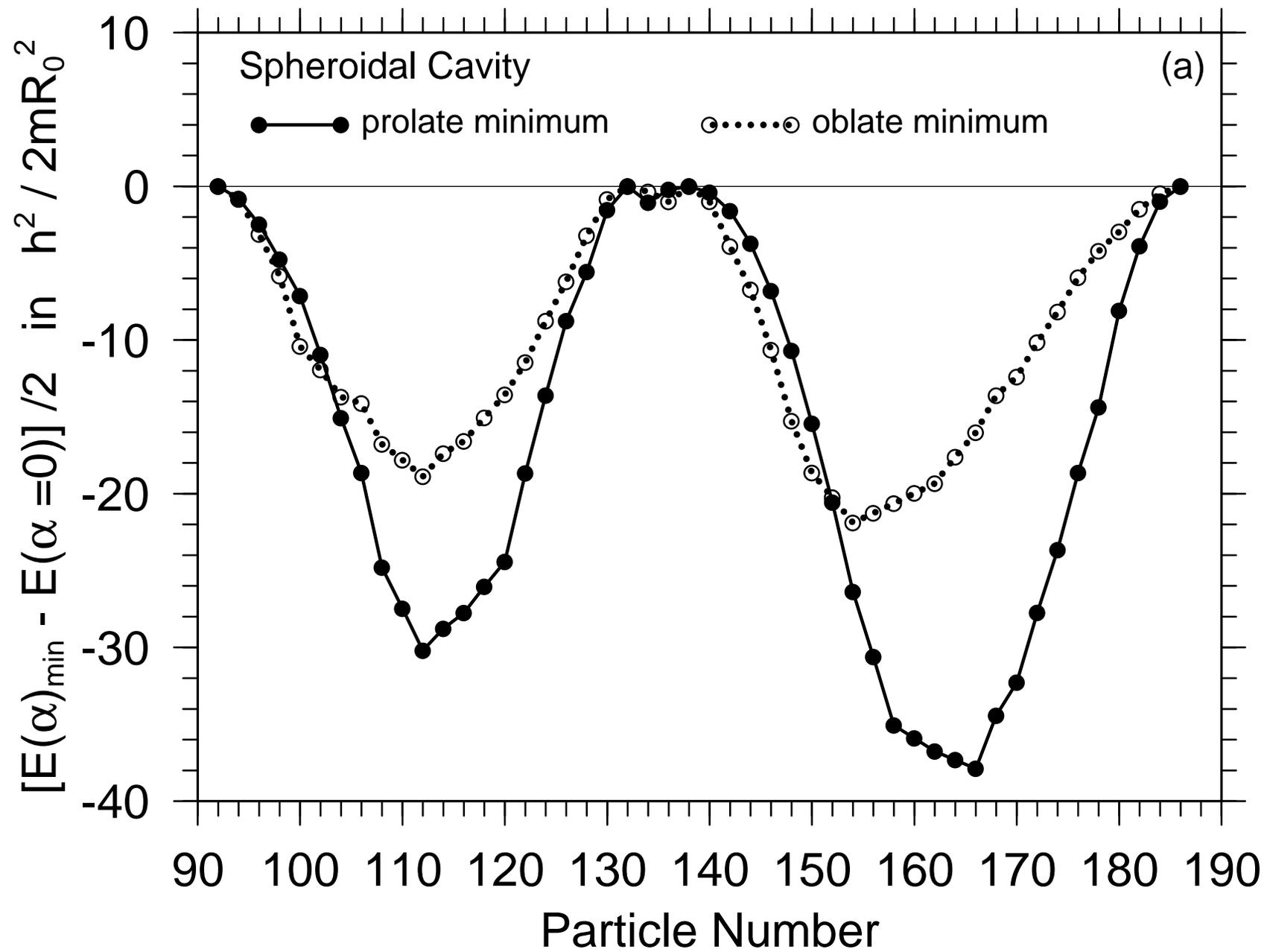
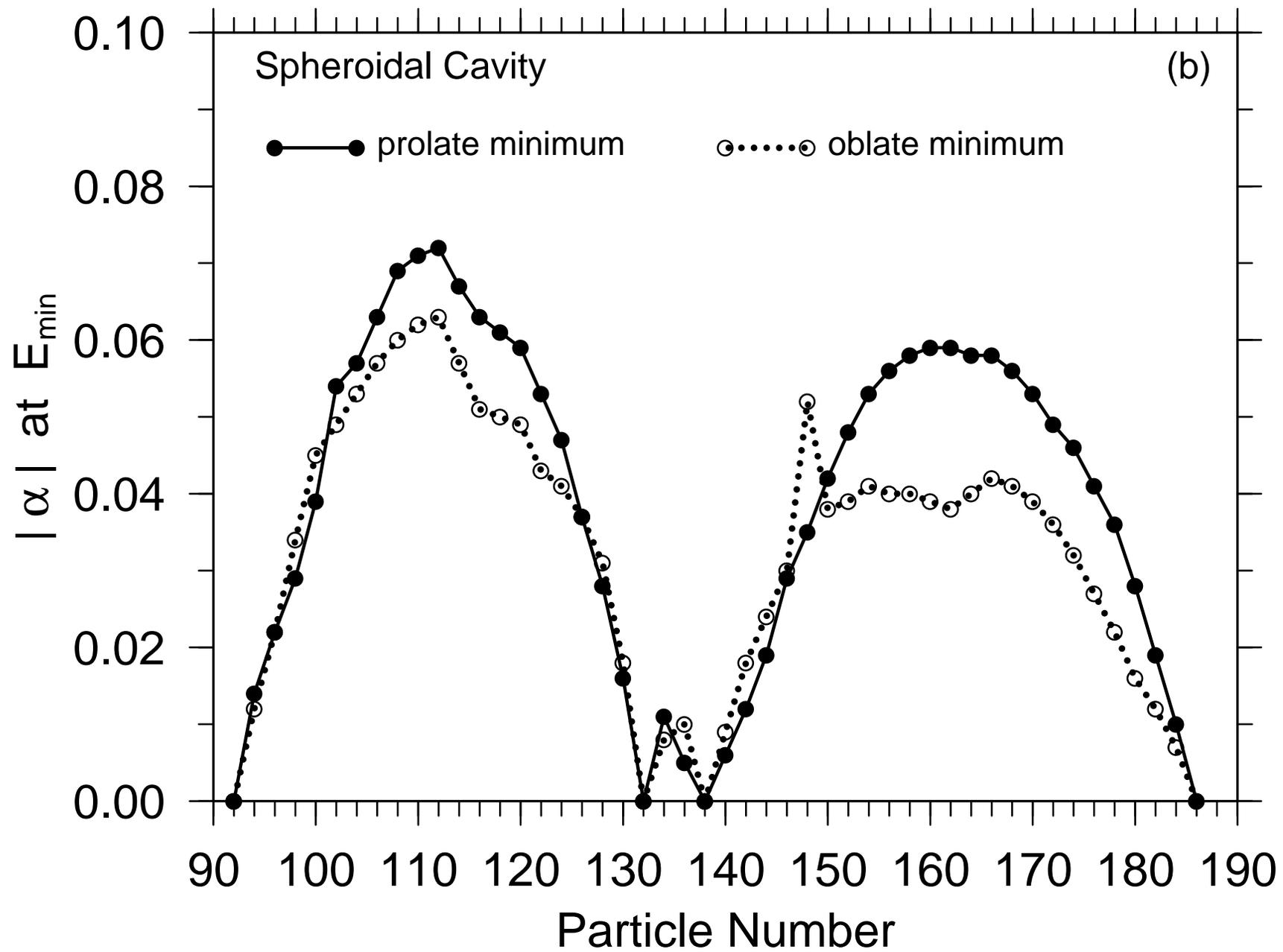
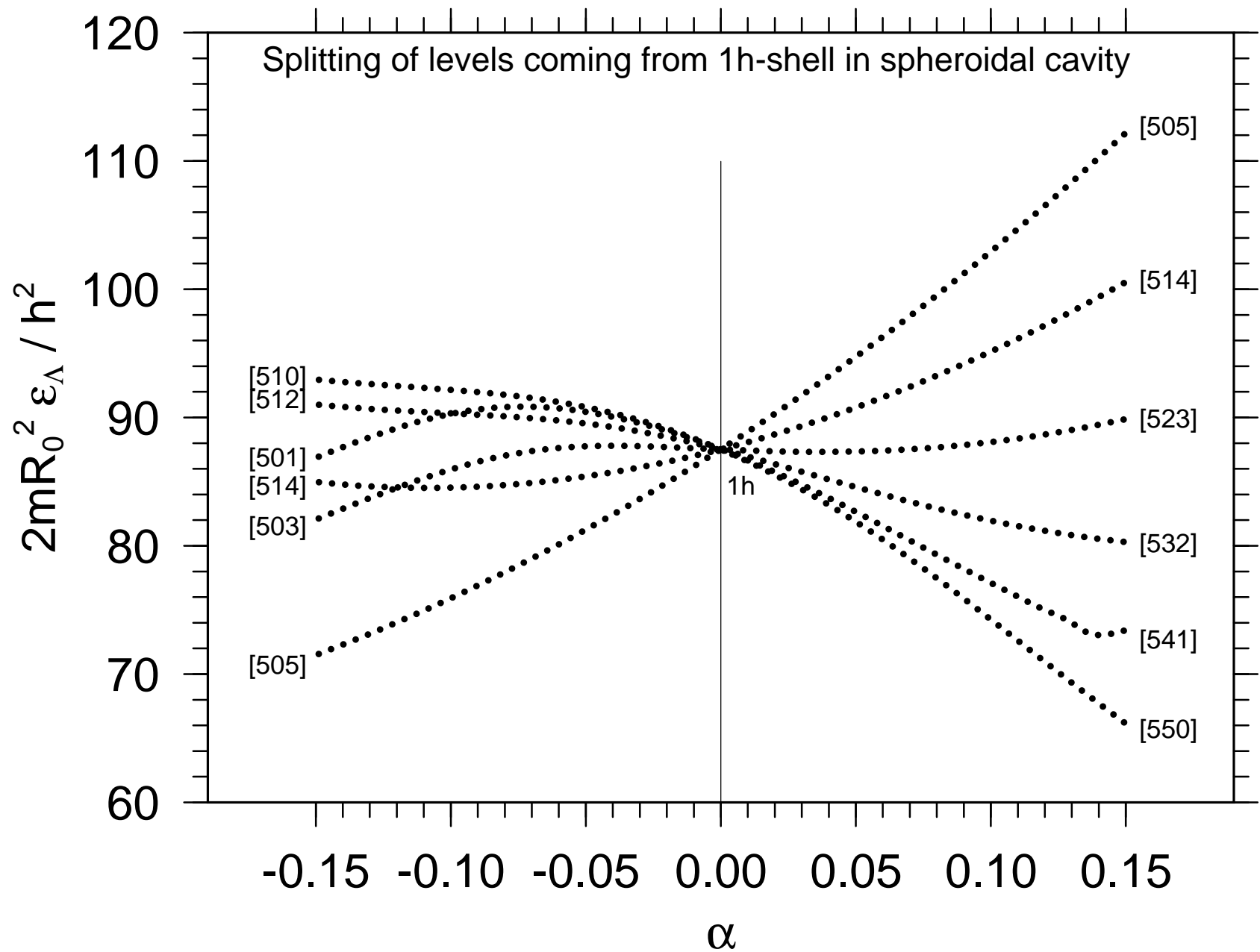
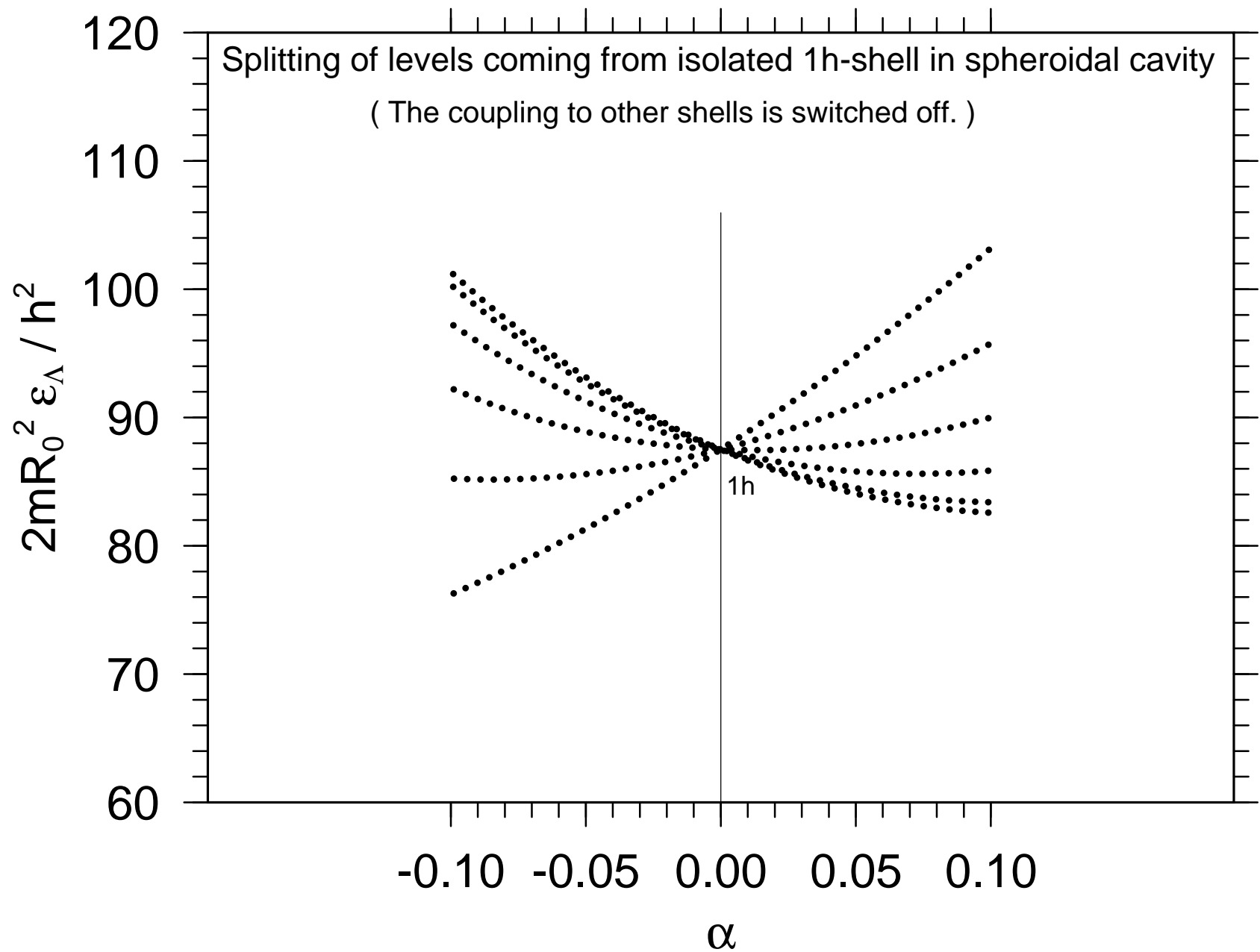
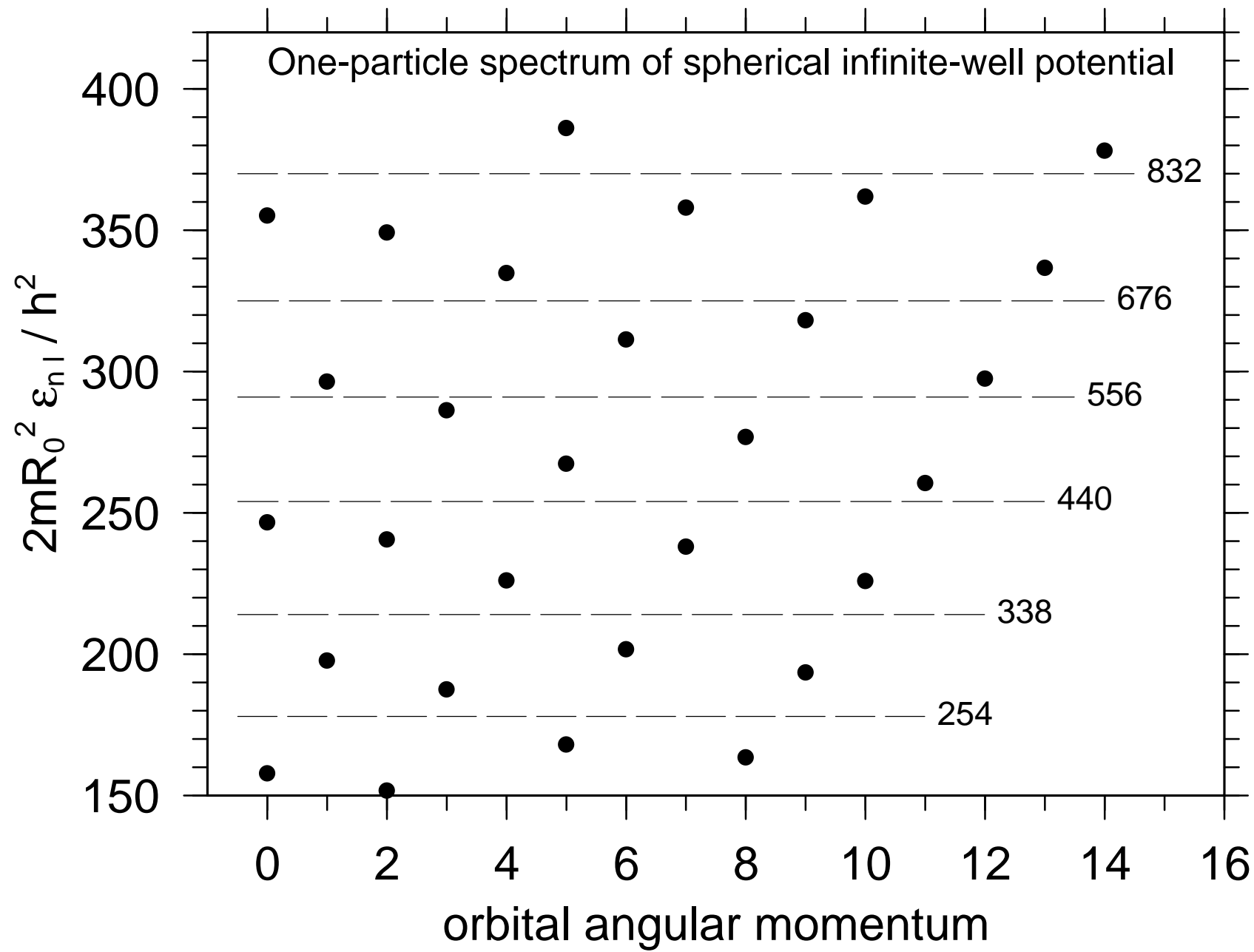
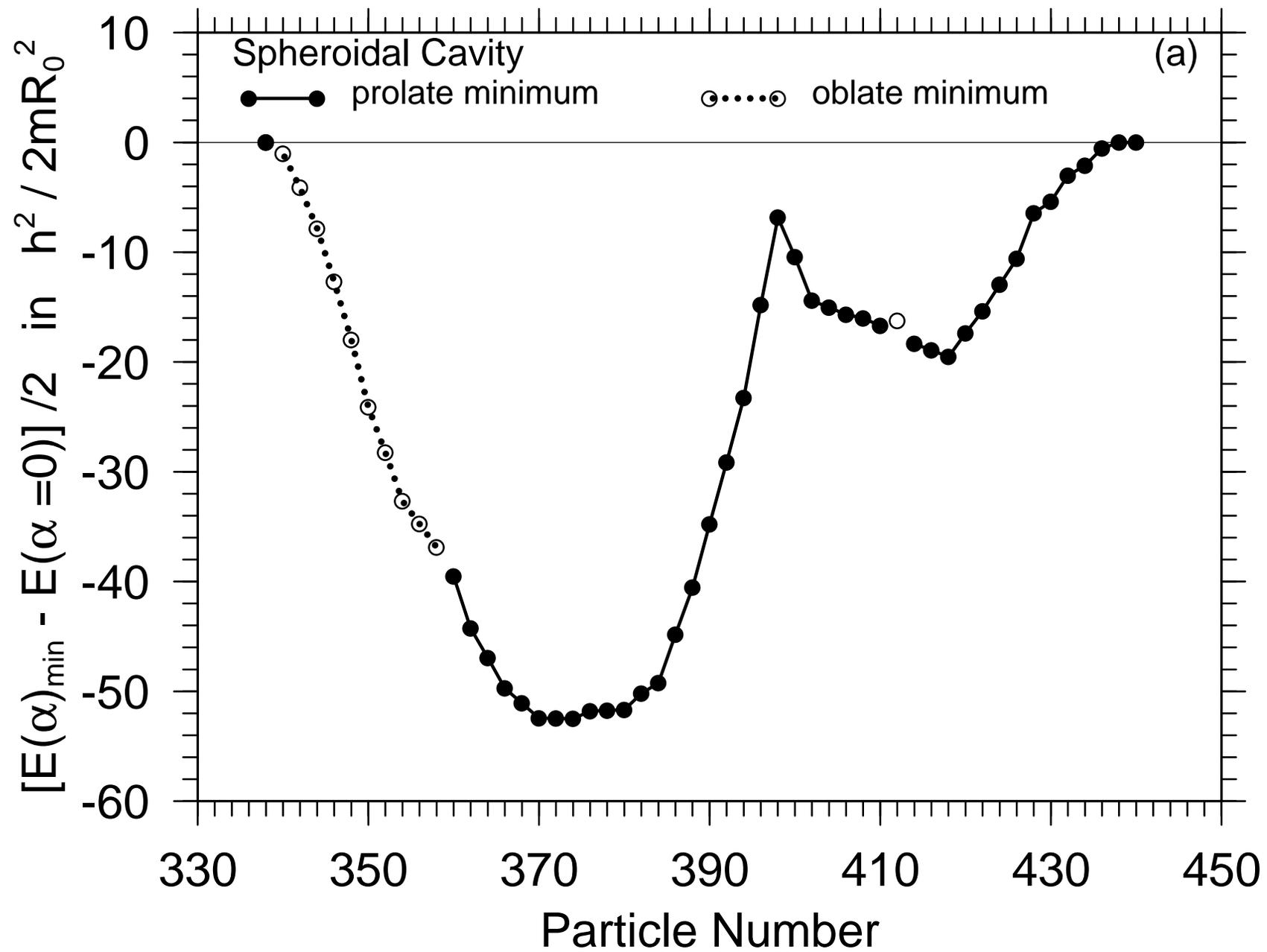
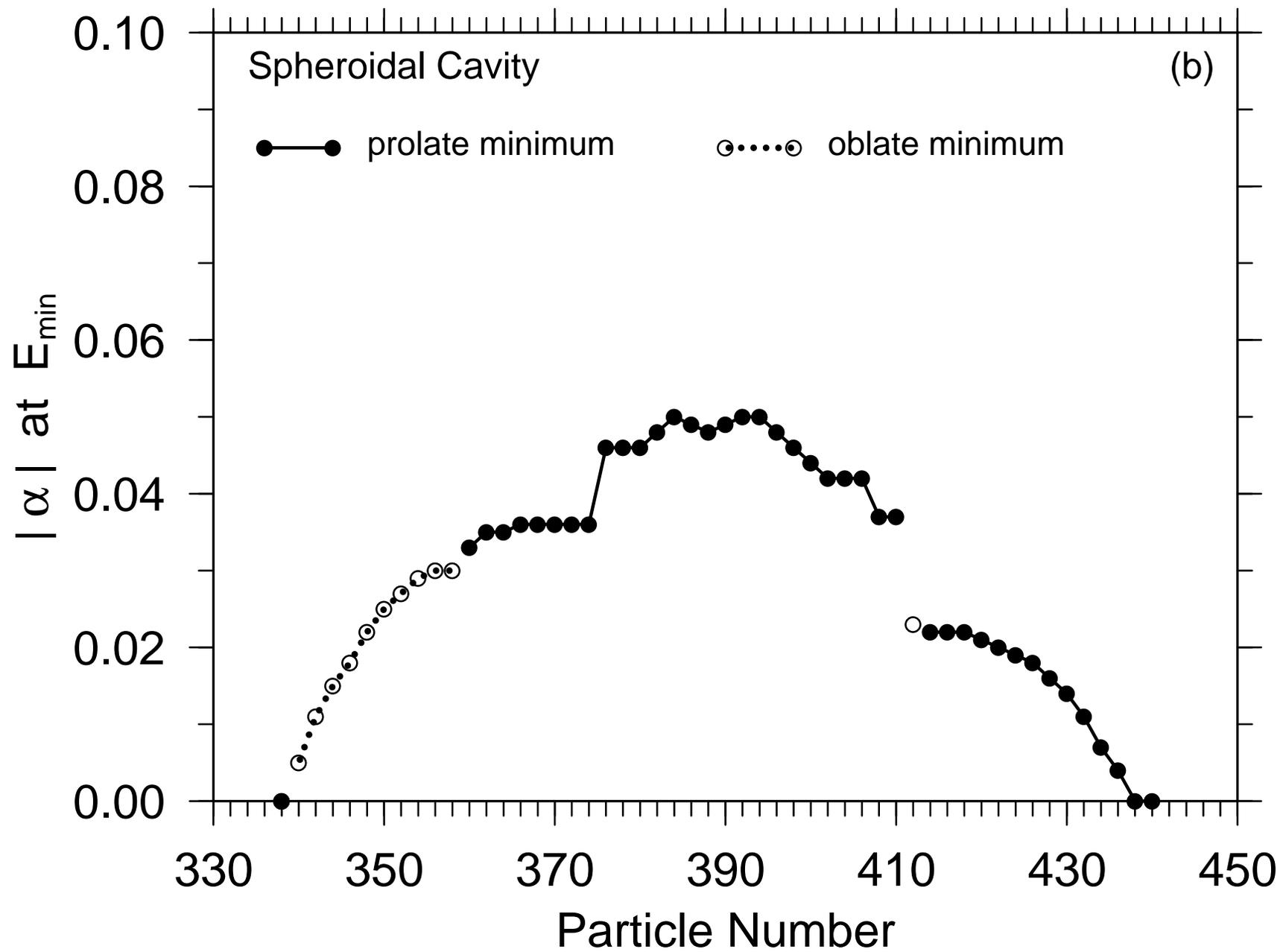
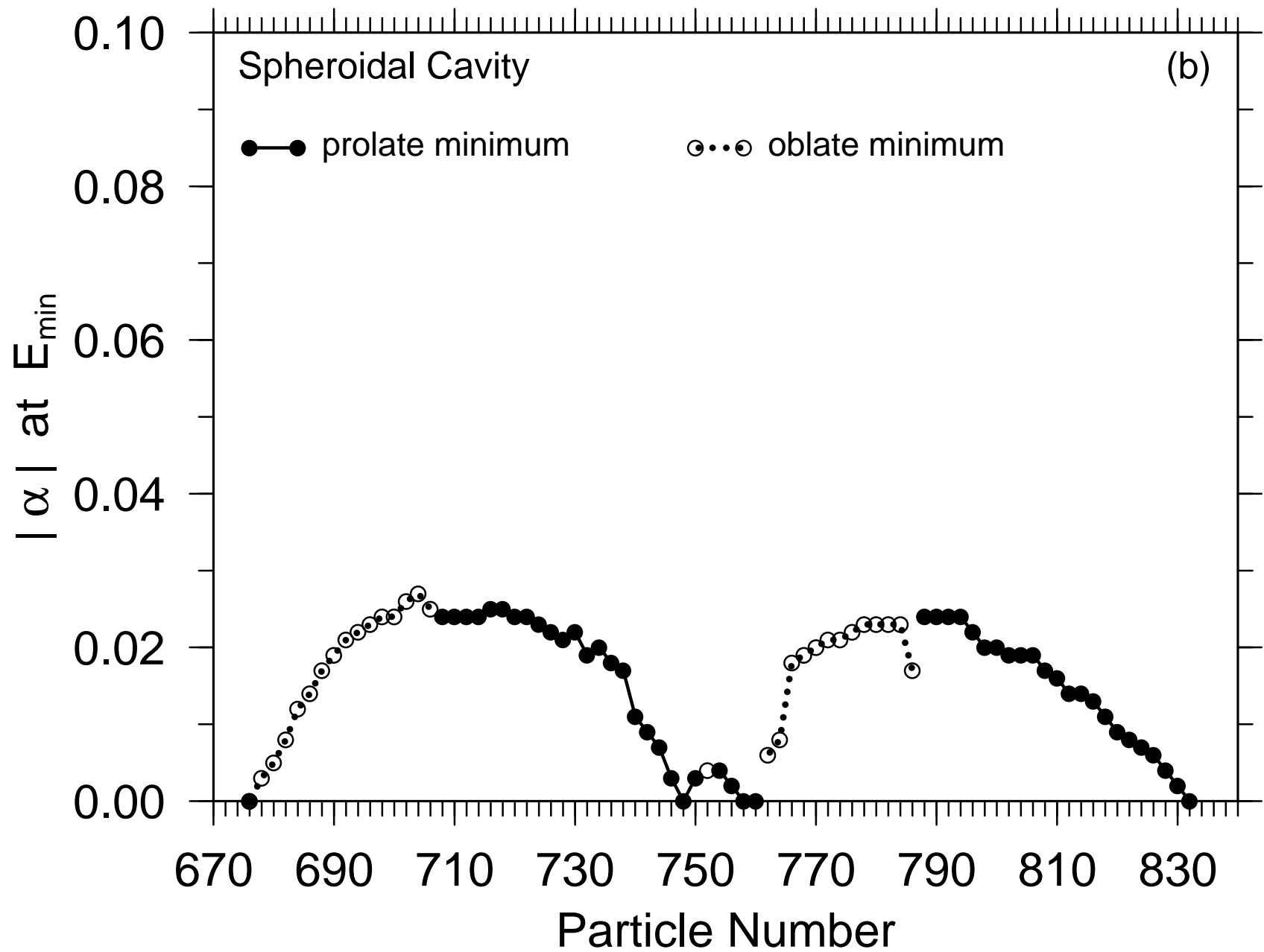
\end{document}